# Subsurface defect damage imaging in PZT ceramics using dual point contact excitation and detection

H. Mahawar[1], K. Agarwal[2], D. K. Prasad[3], F. Melandsø[2], and A. Habib[2]

[1]Department of Electronics Engineering, Indian Institute of Technology, Dhanbad, India

[2]Department of Physics and Technology, UiT The Arctic University of Norway, Norway

[3]Department of Computer Science, UiT The Arctic University of Norway, Norway

## 1. Introduction

The application of piezoelectric materials, such as Lead Zirconate Titanate $(Zr_xTi_{1-x})O_3$ (PZT) is increasing in multiple dynamic industries such as structural health monitoring, wireless energy harvesting devices, measuring blood flow, etc.[1-3] This can be attributed to its diverse unique advantages, such as broad-banded operation frequency, superior electromechanical coupling, low power consumptions, and impedance matching with various substrates.[4,5] In extreme undesirable environmental conditions, this material can get prone to several surface and subsurface defects. These damages hinder with the durability, functioning and ultimately the performance of the material as the operational stress tend to concentrate around these defects. Hence, it is crucial to detect any defect in the PZT material to ensure its proper operation. Owing to its several merits, here we have used an innovative approach namely, dual point excitation and detection technique,[6-8] the major advantage being its ability of wideband excitation and detection without demanding any resonance from the properties of the material for its execution. Though the generated images detected the damage, it also inherently possessed significant amount of noise especially from the subsurface level. In order to get a better and clearer idea of the properties of the defects present in the material, it is important to first alleviate the noise from the images. There are many methods available for denoising of an image.[9] After examining the model of the noise in the images, a systematic denoising schemes have been implemented to produce the best possible outcome for studying these defects in the PZT material.

e-mail: anowarul.habib@uit.no

The main aim of this paper is to denoise the images generated by dual point excitation and detection method for subsurface damage detection. Nonetheless, these denoising schemes can be extended for other noisy images.

## 2. Experiments and Results

A similar experimental setup was discussed earlier by our group.[6] A PZT with a $20 \times 20$ mm$^2$ dimensions and 3 mm thickness was chosen and a calibrated damage was introduced on the surface. As it is practically difficult to create a calculated defect on a subsurface, so that a defect was introduced and then another surface is made on the top of it. Here, we made a surface of 0.5 mm thickness of the same material with a thin layer of ultrasound gel as the coupling media between the sandwich PZT.

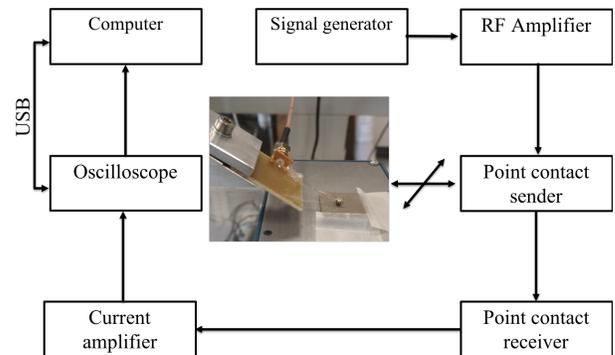

**Fig. 1:** Schematic diagram of point contact excitation and detection method

Fig. 1 represents the schematic diagram of the experimental setup. All these components include an arbitrary function generator for generating the excitation pulse. Later on, the excited signal was amplified with a radio frequency (RF) amplifier and delivered to the steel sphere which was gently in contact with piezoelectric ceramics. A Dirac delta pulse (width 60 ns) form was delivered for excitation. The images of the subsurface defect which contained a significant amount of noise, as the attenuation and refraction from the coupling media and the presence of PZT/coupling media/ PZT contributed to the noise in the images which was detected from the surface of the PZT. The scanning area on the PZT ceramics were set to 10 mm in both directions. Step size were set to 50 μm in x and y directions. Hybrid denoising method; deep neural network based denoiser with adaptive and non-adaptive filtering were incorporated to denoise such kind of noisy images. Filtering involved the use of one frame for each pixel comprising of neighborhood pixels of a specified size to estimate and yield local image averages and deviations.

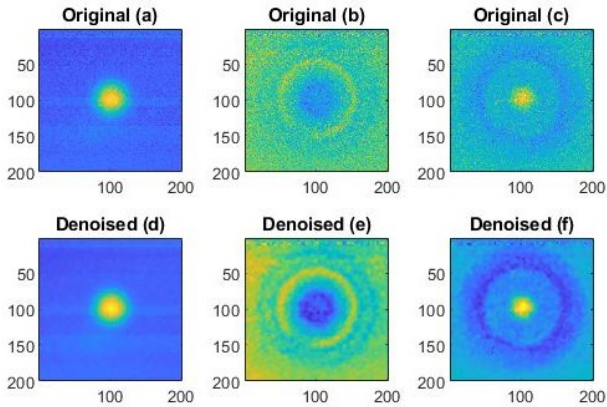

**Fig. 2:** Original and denoised images of ultrasonic wave at three different time frames. Image size of each frame is 10 ×10 mm$^2$.

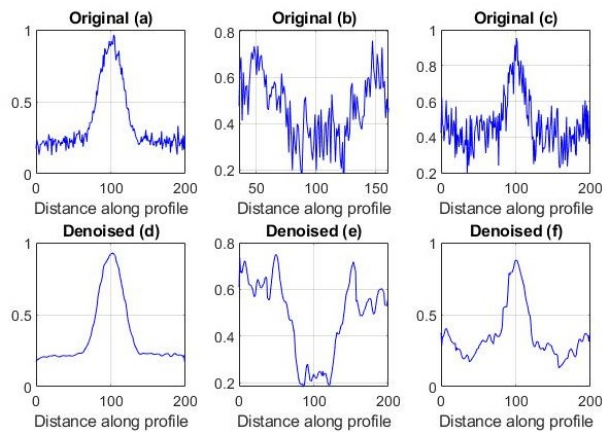

**Fig. 3:** Intensity values along a common line in the images of three different time frames

Fig. 2 shows images which is denoised by a combination of two spatial filtering methods. Due to the absence of noiseless reference image existing denoising metrics are not applicable in this situation.

Hence, we observed the denoising visually by plotting the intensity of the pixels along a horizontal line passing through the center of both the original and denoised images as shown in Fig. 3.

After taking the difference of the intensity values for each pixel between the original and denoised images and adding the square of the result for all pixels we calculated the energy of the noise in the original image that has been removed. Table 1 shows the value of these removed noise.

**Table 1:** Energy of the noise removed after denoising from the images

|  | Image (a) | Image (b) | Image (c) |
|---|---|---|---|
| Energy of the noise removed | 37.08 | 336.24 | 960.64 |

### 3. Conclusion

In order to study effectively, the subsurface defects in a PZT ceramics and from its images the denoising schemes have been examined and a metric for the quantification of the noise is proposed which was previously non-existent. A delta pulse is used for excitation of the acoustic waves in PZT ceramics. The metric aids in calculating the energy of the noise been removed and also to verify the proficiency of the denoising technique been incorporated.